# Low temperature ferromagnetic properties of the diluted magnetic semiconductor $Sb_{2-x}Cr_xTe_3$

J.S. Dyck[1,2], Č. Drašar[3], P. Lošťák[3], C. Uher[1]

[1]Department of Physics, University of Michigan, Ann Arbor, MI 48109, USA
[2]Department of Physics, John Carroll University, University Heights, OH 44118, USA
[3]Faculty of Chemical Technology, University of Pardubice, Čs. Legií Square 565, 532 10 Pardubice, Czech Republic

ABSTRACT

We report on magnetic and electrical transport properties of $Sb_{2-x}Cr_xTe_3$ single crystals with $0 \leq x \leq 0.095$ over temperatures from 2 K to 300 K. A ferromagnetic state develops in these crystals at low temperatures with Curie temperatures that are proportional to $x$ (for $x > 0.014$), attaining a maximum value of 20 K for $x = 0.095$. Hysteresis below $T_C$ for applied field parallel to the c-axis is observed in both magnetization and Hall effect measurements. Magnetic as well as transport data indicate that Cr takes the 3+ ($3d^3$) valence state, substituting for antimony in the host lattice structure, and does not significantly affect the background hole concentration. Analysis of the anomalous Hall effect reveals that skew scattering is responsible for its presence. These results broaden the scope of ferromagnetism in the $V_2$-$VI_3$ diluted magnetic semiconductors (DMS) and in ferromagnetic DMS structures generally.

I. INTRODUCTION

There has been intense research activity on the incorporation of magnetic ions into semiconductor hosts since the discovery of ferromagnetism in the Mn-doped III-V semiconductor GaAs [1]. Exploration of alternative diluted magnetic semiconductor (DMS) structures can broaden the scope of our understanding of the underlying physics of these materials more generally and will promote progress toward realizing applications [2]. So far, ferromagnetism has been reported in a variety of semiconductor hosts where Mn is the magnetic ion including II-VI [3], III-V [1,4], and group IV [5] tetrahedrally-bonded semiconductors. Additionally, studies involving incorporation of alternative ions such as chromium have resulted in stimulated magnetic order. Room temperature ferromagnetism has been reported in several systems including Cr-doped AlN thin films prepared by a variety of techniques [6-8], GaN:Cr single crystals [9], and $Zn_{1-x}Cr_xTe$ thin films [10]; whereas Cr-doped GaAs has a Curie temperature below 30 K [11]. These reports show that chromium also yields a rich range of magnetic behavior depending on growth conditions and host band structure.

In this article, we present results on the low temperature properties of Cr-doped $Sb_2Te_3$. Recently, it was found that single crystal forms of $Sb_{2-x}V_xTe_3$ [12] and $Bi_{2-x}Fe_xTe_3$ [13] display ferromagnetic transitions near 22 K and 12 K, respectively. Interestingly, $Sb_{2-x}Mn_xTe_3$ is paramagnetic down to 2 K [14]. The host matrices for this family of DMS structures are narrow-gap semiconductors ($E_g \sim 0.26$ eV for $Sb_2Te_3$) that belong to the group of tetradymite-type (space group $R\bar{3}m - D_{3d}^5$) layered compounds having the formula $V_2VI_3$ (with V = Sb, Bi and VI = Se, Te). The crystal lattice consists of repeated groups of atomic layers ($VI^{(1)}$-V-$VI^{(2)}$-V-$VI^{(1)}$) oriented perpendicular to the c-axis and separated by a van der Waals gap. Incorporation of chromium into $Sb_2Te_3$ results in ferromagnetism below 20 K, providing a third $V_2$-$VI_3$ ferromagnetic DMS structure.

Taken in the context of all ferromagnetic DMS compounds, this study provides an opportunity to examine the desired magnetic and transport phenomena arising in an alternative system. The anomalous Hall effect (AHE) has been a key property for investigation in DMS materials. Chromium-doped $Sb_2Te_3$ displays a clear AHE and analysis shows that its origin may be different from the more widely studied III-V-based DMS structures. Also, differences between $Sb_{2-x}Cr_xTe_3$ and the other tetradymite structure DMS materials highlight the unusual magnetic properties of this family of compounds.

## II. EXPERIMENT

Single crystals of $Sb_{2-x}Cr_xTe_3$ with nominal x values between 0 and 0.06 were grown using the Bridgman method. The starting polycrystalline materials for growing the single crystals were prepared from 99.999% pure elemental Sb and Te and from $Cr_2Te_3$. The synthesis of $Cr_2Te_3$ was carried out by heating stoichiometric mixtures of 5N purity Te and Cr to 1350 K for 7 days in evacuated quartz ampoules, followed by 10 days at 700 K. This material was powdered and combined with Sb and Te in the ratio corresponding to the stoichiometry $Sb_{2-x}Cr_xTe_3$ (x = 0, 0.02, and 0.06) and synthesis of a polycrystalline product was prepared in evacuated conical quartz ampoules in a horizontal furnace at a temperature of 1073 K for 48 hours. Single crystals were grown from the same charged conical tubes by the Bridgman method after first annealing at 1003 K for 24 hours and then lowering them through a temperature gradient of 400 K/5 cm at a rate of 1.3 mm/h.

This technique yielded single crystals of 5 cm length and 1 cm diameter. As was found previously [15], the concentration of Cr along the crystal growth direction varied significantly for a given nominal $x$ value, with the highest Cr content near the conical tip of the crystal. Specimens for measurements were cut from single crystals at locations progressively closer to the conical tip with a spark erosion machine. Electron microprobe analysis (EMPA) was employed to analyze the stoichiometry, and reported values of chromium concentration, $x$, are taken from the average of 5 readings from a freshly cleaved surface. The measured $x$ values are given in Table I.

Both transport and magnetic property measurements were carried out on the same samples over the temperature range 2 K to 300 K. Magnetic susceptibility and magnetization measurements were made in a Quantum Design SQUID magnetometer equipped with a 5.5 T magnet in field orientations both parallel and perpendicular to the c-axis. Hall effect and electrical resistivity data were collected in the same instrument with the aid of a Linear Research ac bridge with 16 Hz excitation. The current was perpendicular to the c-axis for the transport measurements, and the magnetic field was oriented parallel to the c-axis for the Hall measurements.

III. RESULTS and DISCUSSION

Figure 1 displays the temperature dependence of the in-plane, zero-field electrical resistivity $\rho$ (current $\perp$ c-axis) and Hall coefficient $R_H$ (obtained in a magnetic field of +/- 1 T) of the $Sb_{2-x}Cr_xTe_3$ single crystals. The in-plane electrical resistivity of pure $Sb_2Te_3$ is dominated by hole conduction and has a metallic temperature dependence characteristic of a degenerately doped semiconductor. Hall concentrations, $p = 1/eR_H$ where $e$ is the elementary charge, of $1 \times 10^{20}$ cm$^{-3}$ are typical of undoped Sb$_2$Te$_3$ and are ascribed to the presence of a large number of native antisite defects [16]. As the concentration of Cr in the lattice increases, the electrical resistance increases smoothly over the entire temperature range. The Hall coefficient has a non-monotonic behavior as a function of $x$, with $R_H$ first increasing by about 40% followed by a shallow decrease with increasing $x$. The inset to Fig. 1(a) shows the room temperature Hall concentration as a function of Cr content (filled symbols). This data suggests that the concentration of holes is not strongly affected by the presence of Cr. One possible explanation for the initial suppression of the Hall concentration is that the presence of low concentrations of Cr affect the concentration of native defects [17,18]. The transport data are in agreement with Ref.15 in which the temperature range above 100 K was explored.

At low temperatures, the resistivity data for Cr content greater than $x = 0.014$ display a maximum signaling a paramagnetic-ferromagnetic phase transition. Magnetic semiconductors commonly exhibit a maximum in $\rho$ near their Curie temperature [19], and this can be understood on theoretical grounds in terms of a model that combines scattering by magnetic impurities and magnetic fluctuations [20,21]. Similar features have been observed in other diluted magnetic semiconductors such as Ga$_{1-x}$Mn$_x$As [21], In$_{1-x}$Mn$_x$Sb [4], and Sb$_{2-x}$V$_x$Te$_3$ [12]. Hall coefficient data also reveal an upturn at low temperature occurring at the same temperature that the $\rho$ data display a peak. This behavior is caused by a strong anomalous component in the transverse resistivity, which will be further elucidated below.

Representative magnetization data for temperatures less than 50 K, together with the corresponding zero-field resistivities, are shown in Fig. 2 for $Sb_{2-x}Cr_xTe_3$ with $x =$ 0.047 and $x = 0.095$. The data were collected while cooling in a field of 0.1 T oriented parallel to the c-axis. The magnetization $M$ displays a sharp rise at $T_C$, as expected from the onset of long-range ferromagnetic order. One can see that the resistivity peak $T_\rho$ occurs at the same temperature as the point of inflection in $M(T)$, and this correspondence will serve to determine the Curie temperature in our study. The inset to Fig. 2 shows a plot of $T_C$ as a function of chromium content, $x$. The linear dependence of $T_C$ on $x$ is expected for diluted magnetic semiconductors [22]. Our lowest temperature of 1.8 K was not low enough to determine $T_C$ for two specimens with $x = 0.014$, and these samples did not display measurable hysteresis in the $M(H)$ curves.

From the high field saturation magnetization at 2 K ($M_{sat}$), we have estimated the number of Bohr magnetons ($\mu_B = 9.27 \times 10^{-21}$ emu) per Cr ion in the ferromagnetic state. These values ($M_{sat}/N$ where $N$ is the number of Cr atoms per gram of sample as obtained from the EMPA value of $x$) were calculated to be $3.5\,\mu_B$, $3.10\,\mu_B$, $3.12\,\mu_B$, and $2.98\,\mu_B$ for $x = 0.031, 0.047, 0.069$, and $0.095$, respectively. The relative error in the microprobe results, which is the primary source of error in the determination of the Cr magnetic moment, decreased from >10% for the smallest Cr concentrations to <3% for the highest. Within this error, the Cr magnetic moment is close to $3\,\mu_B$, or to a spin-only value of S = 3/2 per Cr. One expects $3\,\mu_B$ from the $3d^3$ electron configuration for $Cr^{3+}$ if the chromium substitutes for antimony in the lattice and provided the orbital component is quenched. A number of chromium telluride compounds ($CrTe$, $Cr_3Te_4$, $Cr_2Te_3$) are known to display ferromagnetic behavior with saturation moments near $2.0 - 2.7\,\mu_B$ per Cr, though with Curie temperatures between 180 and 340 K [23], differentiating the present results from the magnetic signal due to potential unwanted impurity phases.

Our measurements of $M(T)$ also allow us to obtain the exchange energy $J$ associated with the ferromagnetic state. We can understand the low-temperature limit of $M(T)$ within a standard three-dimensional spin wave model [24] which predicts $M(T) = M_0 - 0.117\,\mu_B\,(k_BT/2SJd^2)^{3/2}$. Here, $M_0$ is the zero temperature magnetization and $d$ is the distance between Cr ions. According to the fit to the data in Fig. 2 (solid lines) at temperatures well below $T_C$, $J$ is computed to be 0.04 and 0.12 meV for the $x = 0.047$ and $x = 0.095$ specimens, respectively, which is comparable to GaMnAs thin films [25] with similar magnetic impurity concentrations. In the three-dimensional Heisenberg model with nearest-neighbor interactions, the Curie temperature is found to be $T_C = 2JS(S+1)z/3 k_B$ where $z$ is the number of nearest-neighbor spins. For $S$ we take 3/2, and we somewhat arbitrarily choose $z = 6$ (cubic lattice of Cr ions) although its magnitude is not well defined in this random system. The open symbols in the inset to Fig. 2 show the excellent agreement to the data from the point of inflection in $M(T)$

Above $T_C$, the magnetic susceptibility $\chi$ has paramagnetic behavior for Cr-doped specimens. Antimony telluride is diamagnetic [26] and we measure a temperature independent value of $\chi = -3.8 \times 10^{-7}$ cm$^3$/g. For temperatures above 100 K, the Sb$_{2-x}$Cr$_x$Te$_3$ data fit well to a Curie-Weiss law of the form $\chi(T) = \dfrac{C}{T - \theta_{CW}} + \chi_0$ where $C$ is the Curie constant, $\theta_{CW}$ is the paramagnetic Curie temperature, and $\chi_0$ is a temperature independent term to account for the diamagnetic host and any Pauli paramagnetism contribution. The fitting parameters are given in Table I. Figure 3 displays the temperature dependence of the quantity $(\chi - \chi_0)^{-1}$. Calculations of the effective Bohr magneton number $p_{eff}$ were made via the equation $C = N p_{eff}^2 \mu_B^2 / 3k_B$ where $N$ is again the number of Cr ions per gram of sample, $\mu_B$ is the Bohr magneton number, and $k_B$ is Boltzmann's constant. Values of $p_{eff} = g\sqrt{J(J+1)}$ are close to the spin only value for $Cr^{3+}$ ($L = 0$, $S = 3/2$) of 3.87 $\mu_B$ taking the Landé g-factor to be $g = 2$, in agreement with low temperature saturation moments that also indicate $S \cong 3/2$. We note that the Curie-Weiss temperatures calculated from the fit are larger than the observed Curie temperatures determined by $M(T)$ and $T_\rho$ data, though the linearly increasing trend with $x$ is qualitatively the same. If we include the data down to temperatures of 50 K in the Curie-Weiss analysis, we obtain a somewhat poorer fit; however, Curie-Weiss temperatures are closer to the respective $T_C$ values and calculated spins per Cr ion are larger by 10 – 20 %. This fact suggests that a fraction of the Cr ions may be forming ordered spin clusters at temperatures above $T_C$ and exhibiting superparamagnetic behavior, much like has been observed in carbon doped (Ga,Mn)P [27].

    The appearance of a ferromagnetic state in $Sb_{2-x}Cr_xTe_3$ was unambiguously confirmed by distinct hysteresis loops in the magnetization as well as Hall effect. It is well known that in ferromagnetic materials the Hall resistivity can be expressed as a sum of two contributions: $\rho_{Hall} = R_0 B + R_S M$, where $R_0$ and $R_S$ are the ordinary and the spontaneous (or anomalous) Hall coefficients respectively, $B$ is the applied magnetic field and $M$ is the magnetization. Figure 4 shows the Hall resistivity and magnetization loops for a sample with $x = 0.095$ at T = 5 K. The observation of the anomalous Hall effect (AHE) including clear hysteresis loops establishes Cr-doped $Sb_2Te_3$ as a homogeneous, ferromagnetic diluted magnetic semiconductor. Anomalous Hall effect measurements are

a valuable compliment to magnetization data observed in SQUID measurements. Unlike those in electrical transport, ferromagnetic characteristics observed in the latter can be caused by ferromagnetic precipitates. Hysteresis in both magnetization and anomalous Hall effect was observed in all samples with $x \geq 0.031$.

Much has been made about the nature of the anomalous contribution to the Hall resistivity in diluted magnetic semiconductors. $R_S$ is conventionally described as the sum of two contributions proportional to $\rho$ and $\rho^2$, known as skew scattering and side-jump scattering, respectively [28]. Figure 5 displays a plot of $R_S$ as a function of $\rho$ for several $Sb_{2-x}Cr_xTe_3$ specimens of varying $x$ at temperatures ranging from 2 K to just below $T_C$. The determination of $R_S$ was made from the zero-field remnance of $\rho_{Hall}$, where the ordinary Hall term is zero, and remnant magnetization determined from SQUID measurements. One can see that in the ferromagnetic state, the material closely obeys the relation $R_S = c\rho$ with $c = 1$ T$^{-1}$. The first reports of the AHE in InMnAs [29] and GaMnAs [1] also reported a linear dependence of $R_S$ upon resistivity (with $c$ typically 1.2 - 1.9 T$^{-1}$ for GaMnAs), strongly suggesting that skew scattering is responsible for its appearance. More recently, other researchers [30] have found that $R_S$ in GaMnAs is proportional to $\rho^2$ rather than $\rho$ for $x < 0.07$. This dependence could be caused by disorder-derived side-jump scattering; however, recent theoretical work of Jungwirth [31, 32] pointed out that scattering is not required to produce the result. In this case, the AHE arises from an anomalous contribution to the Hall conductivity $\sigma_{Hall}$, which in clean samples is not dependent on $\rho$. Taking $\sigma_{Hall} = -\rho_{Hall} / (\rho^2 + \rho_{Hall}^2)$ and letting $\rho_{Hall} \ll \rho$ one obtains $\sigma_{Hall} = -\rho_{Hall}/\rho^2$ -- or $\rho_{Hall}$ proportional to $\rho^2$. The fact that $R_S \sim \rho$ in our material suggests that the AHE is disorder-driven rather than an intrinsic property related to the band structure. This difference from III-V-based DMS structures is somewhat surprising. While the band structures of $Sb_2Te_3$ and GaAs are not similar, the concentration of defects is expected to be roughly the same given similar levels of magnetic ion doping and antisite defects. We can postulate, therefore, that the clean-limit theory of Jungwirth is not applicable for $Sb_{2-x}Cr_xTe_3$.

Although ferromagnetism has already been observed in the tetradymite-type semiconductors (e.g., $Sb_{2-x}V_xTe_3$ [12]), the differences among them illustrate the unusual magnetic properties of this system. The efficiency of the chosen magnetic ion in stimulating ferromagnetic order is remarkably different in the two variants with $T_C/x = 200$ K for Cr and $T_C/x = 1000$ for V. This difference is more striking given the fact that the origin of the ferromagnetism is expected to be of the carrier-mediated RKKY-type [12] and, in both variants $p$ is the same (~ $1 \times 10^{20}$ cm$^{-3}$ independent of $x$) while the spin per ion is 1 for vanadium and 3/2 for chromium. One would expect that a higher spin would favor higher Curie temperature. Electron or x-ray spectroscopic studies could reveal differences in the local structure near the magnetic ions and clarify their exact position within the $Sb_2Te_3$ lattice.

Common to the tetradymite structure diluted magnetic semiconductors is anisotropy in the magnetic properties. A difference lies in the magnitude of this anisotropy. The easy axis for magnetization in these compounds is parallel to the c-axis, and for vanadium-doped $Sb_2Te_3$, the anisotropy field (field where the magnetization parallel to the c-axis was equal to that perpendicular to the c-axis) was much greater than 5 T [12]. A comparison of hysteresis loops of the magnetization at 2 K for $Sb_{2-x}V_xTe_3$ with x = 0.03 and $Sb_{2-x}Cr_xTe_3$ with $x = 0.095$ is shown in Fig. 6. The coercive field $H_C$ is much greater for the vanadium-doped $Sb_2Te_3$ with $H_C$ = 1.2 T than for the Cr-doped $Sb_2Te_3$ with $H_C$ = 0.015 T. Further, while both compounds show an easy axis along the c-axis, the Cr-doped compound reaches a fully saturated state at a much lower field for $B$ perpendicular to the c-axis (about 2 T) as compared to V-doped $Sb_2Te_3$ under the same conditions ($B \gg 5$ T). This fact futher indicates that the mechanism of magnetic order and domain structure may be somewhat different in the two materials.

CONCLUSIONS

Single crystals of $Sb_{2-x}Cr_xTe_3$ for x ≥ 0.031 are ferromagnetic, with Curie temperatures that depend linearly on chromium content, $x$. The ferromagnetic state was confirmed by hysteresis in both magnetization and Hall effect versus magnetic field. The anomalous Hall coefficient is independent of temperature below $T_C$ and most likely arises from skew scattering. Magnetization studies reveal that the easy axis of magnetization is parallel to the c-axis of the structure. Each Cr ion contributes approximately 3 $\mu_B$ to both the paramagnetic signal above 100 K and to the low temperature saturation magnetization. Together with room temperature Hall data indicating a very weak dependence of the apparent hole concentration on x, a magnetic moment of 3 $\mu_B$ per Cr suggests that chromium takes the 3+ ($3d^3$) valence state, substituting for antimony (with formal valence of 3+) in the lattice. These results demonstrate the existence of a third transition metal element (along with vanadium in $Sb_2Te_3$ and iron in $Bi_2Te_3$) that stimulates ferromagnetism in a tetradymite structure narrow-gap semiconductor. Differences among the compounds, however, highlight the unusual magnetic properties of this family of diluted magnetic semiconductors.

ACKNOWLEDGEMENTS

This work was supported by the National Science Foundation grants NSF-INT 0201114 and NSF-DMF 0305221, and by the Ministry of Education of the Czech Republic under the project KONTAKT ME 513.

FIGURE CAPTIONS

Figure 1. (a) Hall coefficient $R_H$ and (b) in-plane electrical resistivity $\rho$ as a function of temperature for $Sb_{2-x}Cr_xTe_3$ single crystals. Current is perpendicular to the c-axis. The Hall concentration at room temperature is shown in the inset.

Figure 2. Temperature dependence of the magnetization $M$ (cooling in B = 1000 G parallel to c-axis) and in-plane resistivity $\rho$ for $Sb_{2-x}Cr_xTe_3$ single crystals with $x = 0.047$ (triangles) and $x = 0.095$ (circles). The solid lines are fits to the spin-wave model. See text for details. The inset shows the Curie temperature $T_C$ as a function of $x$ as determined from the M versus T point of inflection (solid squares) and from three-dimensional spin-wave and Heisenberg models of the low temperature M versus T data.

Figure 3. Inverse magnetic susceptibility versus temperature for $Sb_{2-x}Cr_xTe_3$ single crystals. The lines are linear extrapolations illustrating the ferromagnetic (positive) Curie-Wiess temperatures.

Figure 4. Hysteresis loops in magnetization $M$ and Hall resistivity $\rho_{Hall}$ at a temperature of 5 K for a single crystal of $Sb_{1.905}Cr_{0.095}Te_3$. Magnetic field is oriented parallel to the c-axis.

Figure 5. Anomalous Hall coefficient $R_S$ as a function of electrical resistivity $\rho$ for $Sb_{2-x}Cr_xTe_3$. Data are taken from all samples with $x \geq 0.031$ and at temperatures ranging from

2 K up to the respective Curie temperatures. The dashed line illustrates the relation

$R_S = c\,\rho^1$ which is consistent with AHE due to skew scattering.

Figure 6. Comparison of hysteresis in magnetization with applied magnetic field both parallel and perpendicular to the c-axis for Cr-doped $Sb_2Te_3$ and V-doped $Sb_2Te_3$ taken at 2 K.

Figure 1.

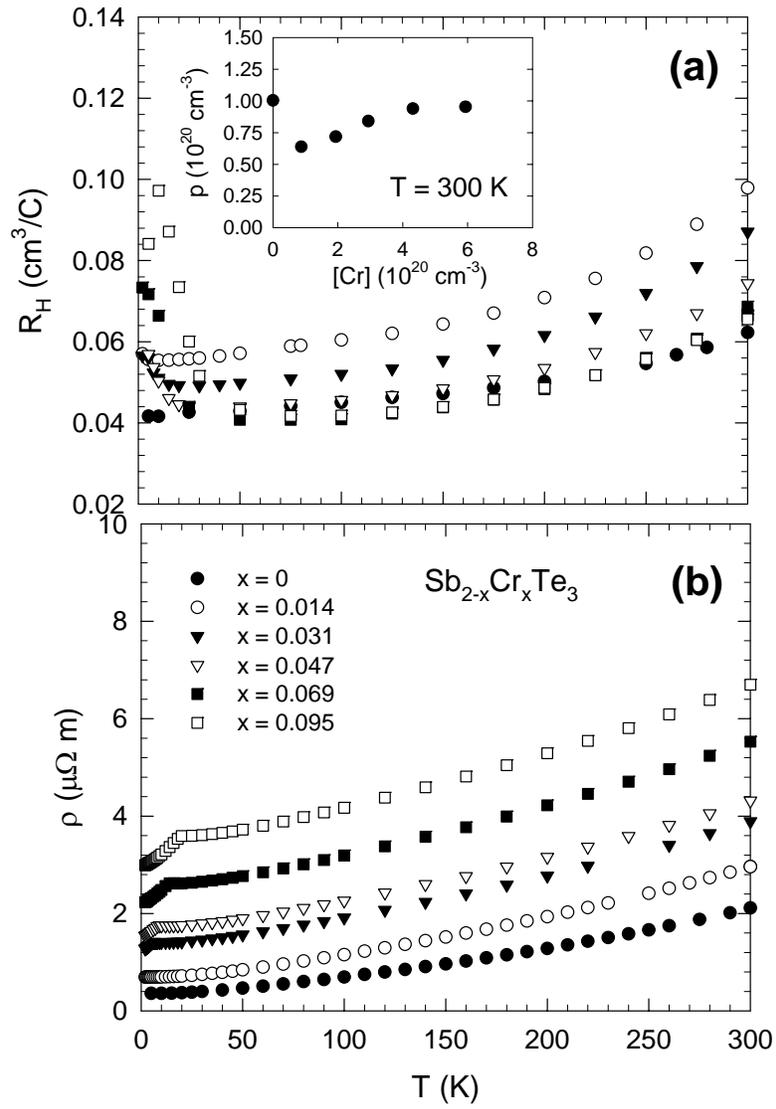

Figure 2.

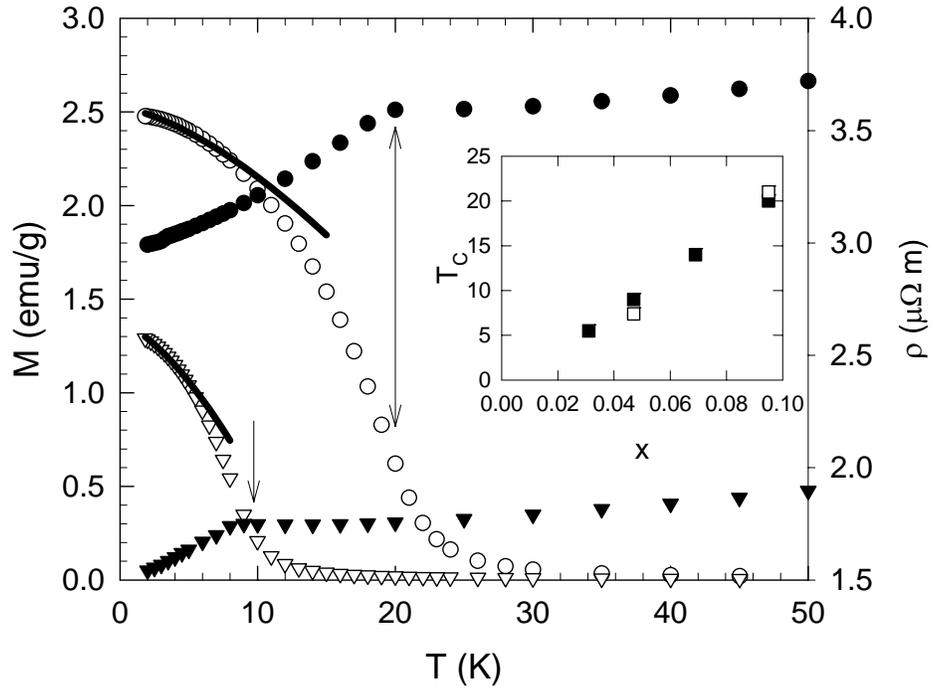

Figure 3.

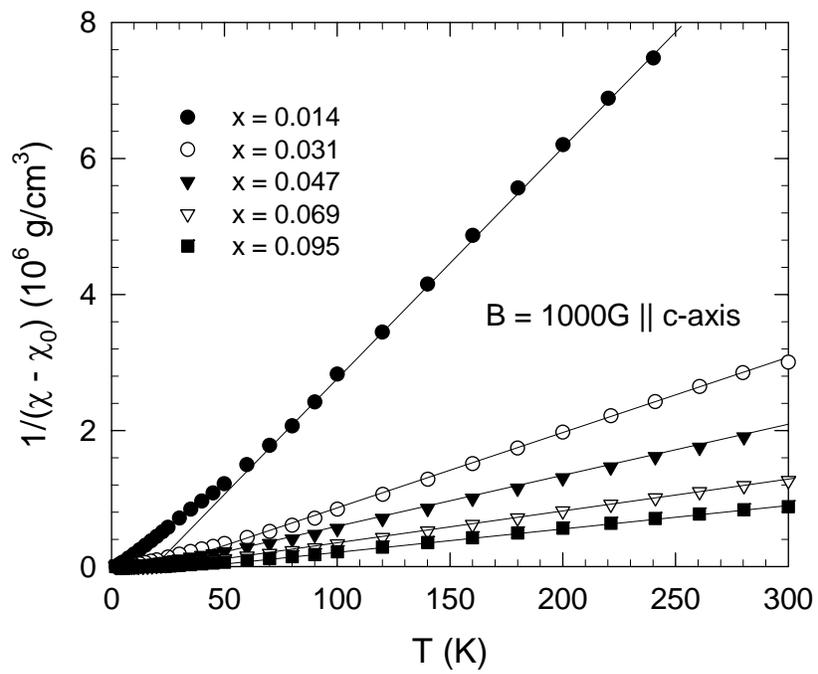

Figure 4.

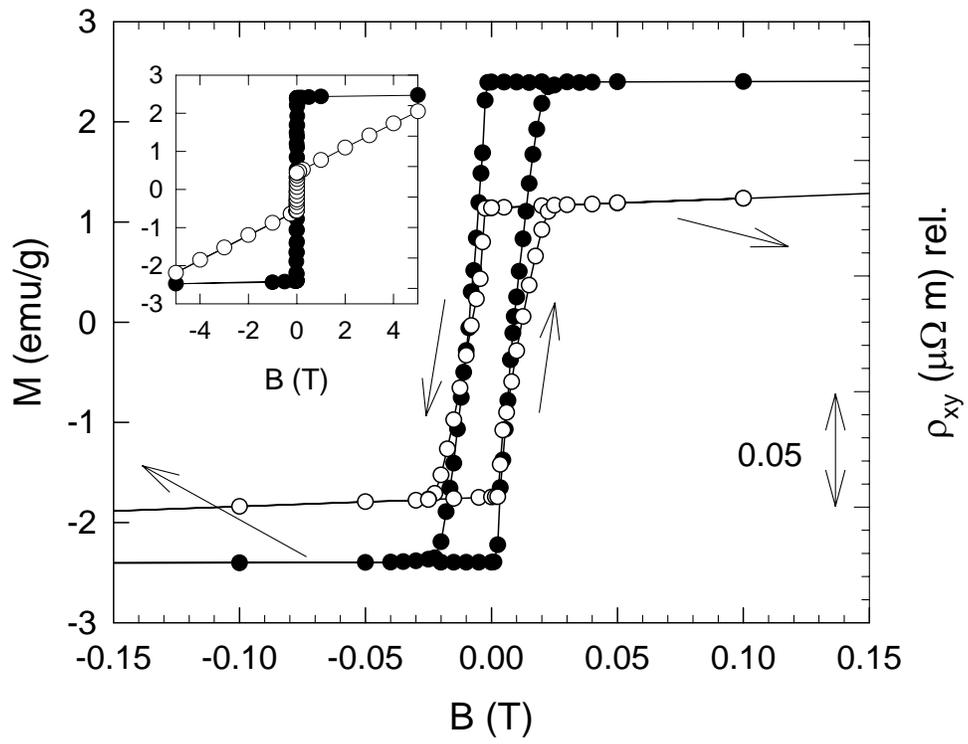

Figure 5.

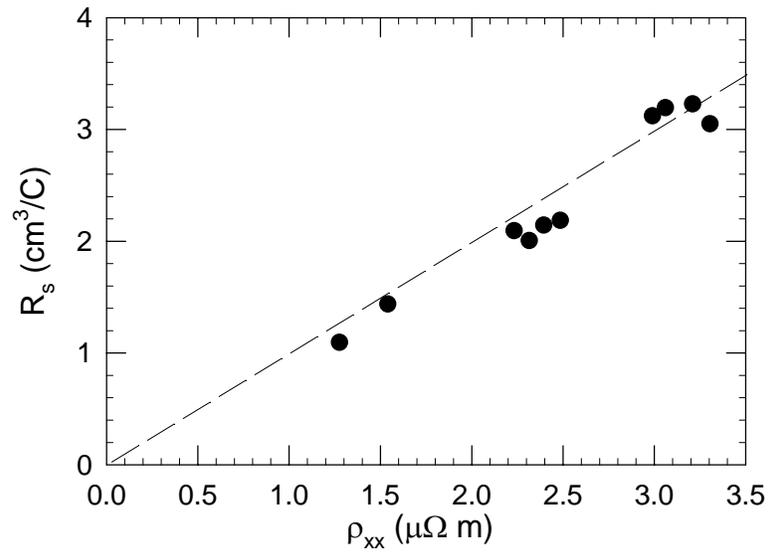

Figure 6.

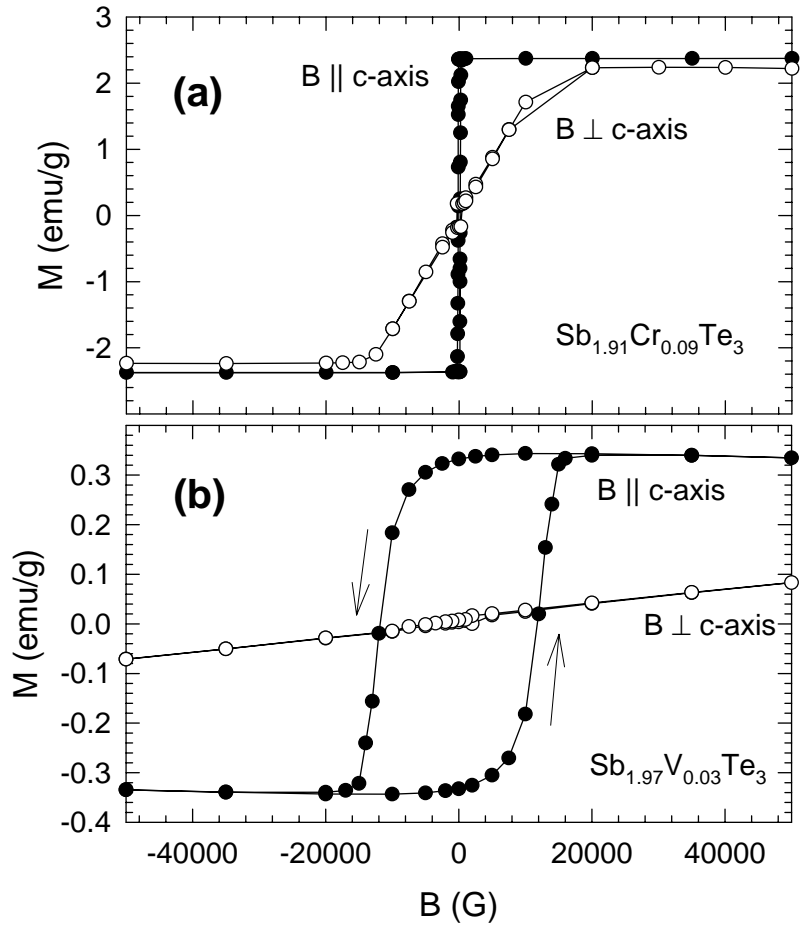

TABLE I. Summary of Curie-Weiss fitting of magnetic susceptibility data over the temperature rage 100 K to 300 K for single crystal $Sb_{2-x}Cr_xTe_3$. Note that the Curie Weiss temperatures $\theta_{CW}$ are higher than TC determined from the point of inflection in $M(T)$

| $x$ (EMPA) | $C$ [cm$^3$K/g] | $\theta_{CW}$ [K] | $\chi_0$ [cm$^3$/g] | $p_{eff}$/Cr | $S$/Cr |
|---|---|---|---|---|---|
| 0.014 ± 0.001 | 3.0062e-5 | 15.1599 | -3.1400e-7 | 3.281 | 1.215 |
| 0.031 ± 0.014 | 8.9440e-5 | 24.9493 | -1.9135e-7 | 3.803 | 1.466 |
| 0.047 ± 0.003 | 1.3486e-4 | 24.7238 | 1.8279e-8 | 3.794 | 1.462 |
| 0.069 ± 0.001 | 2.0959e-4 | 30.6933 | -1.7882e-7 | 3.903 | 1.514 |
| 0.095 ± 0.001 | 2.9090e-4 | 37.110 | -2.3603e-7 | 3.918 | 1.522 |